# Polarization diversity close to the optical bound states in the continuum

Weimin Ye[1*], Yang Gao[3], and Jianlong Liu[2†]

1. College of Advanced Interdisciplinary Studies, National University of Defense Technology, Changsha, 410073, China

2. Department of Physics, Harbin Institute of Technology, Harbin 150001, China

3. College of Electronic Engineering, Heilongjiang University, Harbin 150080, China

*wmye72@126.com

†liujl@hit.edu.cn

**Abstract**

Bound states in the continuum (BICs) realized in two-dimensional (2D) photonic crystal slabs (PhCSs) have attracted considerable attentions, owing to the advantages in the fabrications and on-chip applications. The polarization vortex centered at BICs in the momentum space displayed by the linear polarization vectors of far-field radiation is a striking property of BICs. Here, by constructing BICs at K point in photonic graphenes, we theoretically demonstrate that the far-field polarization states close to BICs can exhibit remarkable polarization diversity. Interestingly, *C*-points (circular polarizations) close to BIC at K point are observed. Thus, we propose that, along a closed loop enclosing BICs in the momentum space, the trajectory of the far-field polarization states on the shell of the Poincaré sphere could provide a general and straightforward means to characterize the topological natures of BICs. The topological charge carried by the BIC can be defined by the half of the winding number of the trajectory around the $S_3$ axis, which is an integer and half integer for the non-degenerate and double-degenerate BICs at K point, respectively. Our findings could open a gateway towards the applications of BICs in lasers with full polarizations and the generation and manipulation of vector beams of light.

Optical bound states in the continuum (BICs) [1] are embedded and infinite-lifetime eigenstates existing in continuous spectral range spanned by the propagating modes. BICs are mainly achieved by the symmetry incompatibility with the free-space radiations or the destructive interferences among the different radiation channels in the far field [2]. The former is sensitively dependent on the geometric symmetry of the photonic structures and is called symmetry-protected BIC. It was observed in one-dimensional (1D) waveguide arrays [3] and two-dimensional (2D) photonic crystal slabs (PhCSs) at the Γ point [4]. The latter could be efficiently constructed through tuning parameters of the photonic structures [1], such as 1D waveguiding structures containing anisotropic birefringent materials [5], coupled-waveguide arrays [6], dielectric gratings [7, 8, 9], 2D PhCSs [10, 11, 12] and arrays of dielectric spheres [13]. Meanwhile, in compact systems, a single layered sphere coated with zero-epsilon meta-materials [14] and a single sub-wavelength high-index-dielectric cylindrical resonator [15, 16] have also been demonstrated to support BIC and quasi-BIC by tuning parameters, respectively.

Recently, BICs in 2D PhCSs have gained increasing attentions due to their advantages in the designs, fabrications and on-chip applications. Different from guided modes in PhCSs, the optical BICs reside within the light cone of the surrounding medium and thus are compatible with free-space radiation, which facilitates experimental observations of BICs. In the previous studies, Off-Γ BICs in both passive and active PhCSs have been demonstrated by the disappearance of Fano features in the reflectivity spectrum [10] and the vanishing photocurrent [11], respectively. Double-degenerate BICs at Γ point supported by a square-lattice PhCS were also exploited in optically pumped BIC lasers [12]. Furthermore, when the far fields are linear polarizations, BICs have been verified to be momentum-space vortices centers exhibited by the far-field polarization vectors [17]. This topological property opens new applications for BICs in on-chip creation and manipulation of vector beams and polarization vortices [9, 18, 19]. However, it fails to characterize degenerate BICs because the far-field polarization directions of any degenerate states in the continuum are indeterminate and spontaneously present as a vortex center in the momentum space [19]. Thus, the previously observed polarization vortices were limited to the linear polarizations and non-degenerate BICs in the interior region of the first Brillouin zone (FBZ) whose topological charges were integers [9,17-19]. In addition, the reported theoretical models based on numerical methods [9-12, 20] have relatively little insight on the mechanism underlying the realization of BICs.

In this letter, focusing on the far-field polarization states close to BICs, we study **K**-point BICs (BICs at **K** point) supported by the honeycomb-lattices PhCSs called as photonic graphenes. They are direct analogue to graphene, and could introduce two degrees of freedom, pseudospins [21, 22] and valleys [23, 24] to the photons, which are widely used in 2D topological photonics [25]. We theoretically demonstrate that the far fields of the PhCSs supporting BICs could exhibit remarkable polarization diversity. Circular polarizations, linear and elliptical polarizations with variant orientations and ellipticities can appear in the vicinity of the BICs. Therefore, instead of the linear polarization vortices centered at BICs, we propose to use the trajectories of the far-field polarization states on the Poincaré sphere to characterize the BICs with differently topological natures. Furthermore, considering the couplings between propagation modes inside the PhCSs and the free-space radiations, an analytical model is developed to describe BICs realized by the destructive interferences, which could provide great insight on the realization of BICs and the polarization diversity close to them.

It is worth noting that our findings involving the polarization diversity are inconsistent with the reported consequence [10,17] that when the slab was invariant under symmetry operator $C_2^z T$ ($C_2^z$ means $180^0$ rotation around $z$ axis, and $T$ means the time reversal operator), the far field of its eigenmode was linearly polarized. Considering any eigenmode in the continuum $\boldsymbol{u_k}$ (that is a leaky mode with a complex frequency $\omega$ and Bloch wave vector $\boldsymbol{k}$) of the slab, the electromagnetic field inside the semi-infinite cover (substrate) is an outgoing wave with a nonzero power flow along the outwardly normal line of the slab. Due to the $C_2^z T$ symmetry of the slab, the new mode $\boldsymbol{u'_k}$ obtained by the operator $C_2^z T$ on the mode $\boldsymbol{u_k}$ is also the eigenmode of the slab with the frequency $\omega^*$. On any $x$-$y$ plane outside the slab, the spatial average of $z$-directional power flows of the eigenmodes $\boldsymbol{u_k}$ and $\boldsymbol{u'_k}$ are opposite and nonzero. It means the new mode $\boldsymbol{u'_k}$ must be an incoming wave in the cover (substrate). Therefore, the $C_2^z T$ symmetry of the slab cannot guarantee that two modes $\boldsymbol{u_k}$ and $\boldsymbol{u'_k}$ must differ at most by a phase factor. As a result, the consequence derived in Ref. [10, 17] was incorrect (for details see the Supplementary Material Note 1).

Figure 1(a) shows our considered 2D PhCS composed of a honeycomb array (a lattice constant $a$) of cylindrical holes (identical radii $r$) etched in a free-standing dielectric slab (a thickness $h$ and a refractive index $n$=2.02 corresponding to $Si_3N_4$ [10]). Due to the $z$-mirror symmetry (invariance under the operation $\sigma_z$ changing $z$ to $-z$), the eigenmodes of the PhCS could be divided into TM-like (defined by $\sigma_z$=−1) and TE-like ($\sigma_z$=1) modes, respectively [26]. Figure 1(b) depicts the FBZ of a honeycomb lattice.

At its corners, three equivalent **K** points, denoted by $\mathbf{K}_j$, $j$=1, 2, 3, provide six free-space radiation channels. They are the TM and TE polarized plane waves with the in-plane wave vector equal to $\mathbf{K}_j$. Distinct from Γ-point BICs, none of **K**-point BICs is symmetry-protected because three **K** points are equivalent. At the fixed Bloch wave vector $\mathbf{K}_1$, owing to the *y*-mirror symmetry, the TM and TE polarized plane waves are only symmetry compatibility with even (defined by $\sigma_y$=1 in the Letter) and odd ($\sigma_y$=−1) modes of the PhCS, respectively. Figures 1(c) and 1(d) respectively show the distributions of the calculated reflectivity spectra by the rigorous coupled-wave analysis software S4 [27], when TM and TE polarized plane waves incident onto the PhCSs with the fixed radii *r*=0.15*a* and different thickness *h*. From the disappearances of Fano features in the reflectivity, we obtain that the PhCSs with the thickness and the normalized frequency (*h*, $\omega a/2\pi c$) equal to (1.27*a*, 0.7062) and (0.9074*a*, 0.7998) support non-degenerate even and odd BICs, respectively. Meanwhile, their eigen fields exhibit different *z*-mirror symmetries. The non-degenerate even and odd BIC are TM- and TE-like modes, respectively.

To look into the different topological natures of the BICs, we study the distributions of the far-field polarization states of eigenmodes supported by the PhCSs in the momentum space near the BIC (For details of the method used to obtain the polarization states, see the Supplementary Material Note 2). Since $\mathbf{K}_1$ is a high-symmetry point in the FBZ, the TM-like non-degenerate even BICs is located at the top of one band of the PhCS. Figure 2(a) displays the far-field polarization states of its eigenmodes on the band with Bloch wave *k* in the vicinity of $\mathbf{K}_1$ point. Compatible with the even symmetry ($\sigma_y$=1) of the BIC, the far fields of the eigenmodes with Bloch wave vector along the *x* direction are TM polarizations (shown as a short line in the *x* direction). Very close to $\mathbf{K}_1$ point, the far fields are nearly linear and radially oriented polarizations relative to $\mathbf{K}_1$ point [Fig. 2(a)], which approximately exhibits a polarization vortex centered at the BIC. However, the TE-like non-degenerate odd BIC is located at the bottom of one band of the PhCS. In the vicinity of $\mathbf{K}_1$ point, the far-field polarization states of its eigenmodes on the band exhibit the diversity [Fig. 2(b)], which cannot be simply described by the orientation angles. Remarkably, when Bloch wave number $k_y$ changes from -0.032*K* to 0.032*K* with the fixed $k_x$ equal to -1.022*K*, the far-field polarization states change from right-handedly circular to left-handedly circular polarizations without rotating their directions. Note that in Fig. 2(b) along the boundary between the right- and left-handed ellipses, denoted by a dash line, the far fields of eigenmodes are the linear polarizations whose predominant component of electric field is TM polarization. The point of intersection between the dash line and the line with Bloch wave number $k_y$=0 in the momentum space is just $\mathbf{K}_1$ point (for details see Fig. s1 in Supplementary Material Note 3). Owing to the *y*-mirror symmetry of the PhCS, the polarization direction of the far field at $\mathbf{K}_1$ point is undefined. Thus, the non-degenerate odd eigenmode at $\mathbf{K}_1$ point becomes a BIC with an infinite *Q* value [17]. Moreover, the PhCS shown in Fig. 1(a) with *r*=0.143*a* and *h*=1.036*a* supports a TE-like double-degenerate BICs at the **K** point. Close to the BICs, two bands of the PhCS linearly cross each other and form the Dirac cone dispersion. Figure 2(c) displays the far-field polarization states of eigenmodes in the lower band of the Dirac cone. Due to the *y*-mirror symmetry of the PhCS and the undefined polarization states of double-degenerate modes at $\mathbf{K}_1$ point, the far fields of eigenmodes with Bloch wave number $k_y$=0 are TM and TE polarizations on the left and right sides of $\mathbf{K}_1$ point, respectively. Figure 2(d) presents the similar results of eigenmodes in the upper band. The diverging *Q* value of the eigenmode at $\mathbf{K}_1$ point demonstrates the mode is a BIC. [For details of band structures and dependences of quality factors (*Q*) & eigen-frequencies over Bloch wave vector close to BICs, see Figs. s2(a)-s2(g) in Supplemental Material Note 4]

Considering the existence of the polarization diversity of eigenmodes in the vicinity of BICs, we record the trajectory of the varying far-field polarization states on the shell of the Poincaré sphere, when Bloch wave vector moves along an anticlockwise closed loop enclosing the BICs in the momentum space. For a pure polarization state with the orientation angle *ψ* and ellipticity angle *χ*, the longitude *φ* and the latitude *θ* of its corresponding point on the shell of Poincaré sphere are equal to 2*ψ* and 2*χ*, respectively. Note that the circular polarization is located at the north and south poles with the undefined longitude or orientation angle. In the momentum space, the point corresponding to the wave vector of the circularly-polarized eigenmode is called as *C*-point [28]. Thus, limited to the closed loops *L* enclosing BIC in the momentum space without containing any *C*-points, the topological charge $q_B$ carried by the BIC can be defined as the half of the winding number $n_w$ (an integer) of the corresponding closed trajectory *G*(*L*) around the $S_3$ axis. That is,

$$q_B = \frac{1}{4\pi} \oint_{G(L)} d\varphi = \frac{n_w}{2}. \tag{1}$$

It is obvious that when the far field is linearly polarized, the topological charge defined in Eq. (1) is consistent with that of the linear polarization vortex [17, 29, 30]. Meanwhile, just limited to the closed loops *L* enclosing one *C*-point without containing any other polarization singularities (e.g. *C*-points or BICs), the topological charge $q_C$ of *C*-point [28] can also be given by Eq. (1). Therefore, the winding number $n_w$ of the closed trajectory on the shell of the Poincaré sphere is the double of the sum of the topological charges of the *C*-points and BICs enclosed by the corresponding closed loop in the momentum space.

For the PhCSs supporting the non-degenerate even BIC [Fig. 2(a)] and the double-degenerate BICs [Figs. 2(c), 2(d)], no *C*-points exist in the vicinity of these two BICs. The topological charges carried by them are just equal to the half of the winding numbers of the closed trajectory on the Poincaré sphere [Eq. (1)]. Without loss of generality, we choose the anticlockwise circle centered at **K**$_1$ point with the radius 0.02*K* in the momentum space [Fig. 3(a)] to record the varying polarization states. Figures 3(b)-3(d) show the trajectories of the far-field polarization states of eigenmodes supported by the PhCSs along the circle enclosing the BICs. The winding numbers of the three trajectories are equal to 2, 1, 1. Thus, the topological charges carried by the non-degenerate even BIC and double-degenerate BICs are 1 and 1/2, respectively. While, for the PhCS supporting the non-degenerate odd BIC [Fig. 2(b)], there are two *C*-points located at ($k_x$, $k_y$) equal to (-1.022*K*, -0.032*K*) and (-1.022*K*, 0.032*K*) close to **K**$_1$ point. We select three anticlockwise circles $L_1$, $L_2$, $L_3$ [Fig. 4(a)] enclosing the BIC with the radii $R_k$ equal to 0.016*K*, 0.033*K*, 0.056*K* and the centers ($k_x$, $k_y$) equal to (-*K*, 0), (-1.01*K*, 0.015*K*), (-*K*, 0), respectively. Thus, the amount of *C*-points enclosed by circles $L_1$, $L_2$, $L_3$ is zero, one, two. Figures 4(b)-4(d) show the winding numbers of the trajectories corresponding to the circles $L_1$, $L_2$, $L_3$ are equal to -2, -1, 0. It means the topological charge carried by the BIC and the *C*-point is -1 and 1/2, respectively. [For the projected trajectories (letting $S_3$=0) of three-dimensional trajectories in Figs. 3(b)-3(d) and 4(b)-4(d) on the $S_1$-$S_2$ plane, see Figs. s3 and s4 in Supplemental Material Note 5].

Focusing on the energy conservation and the field inside the slab, we utilize the general scattering-matrix model [31] to investigate the realizations of BICs by the destructive interferences and the polarization diversity close to them. Different from the analytical treatment of Ref. [20], the transversal field inside the slab $|v\rangle = \begin{bmatrix} H_x & H_y & -E_y & E_x \end{bmatrix}^T$ is approximately expressed as $|v\rangle = \sum_{m=1}^{N}\left[\rho_m^{(+)}e^{i\beta_m(z-h/2)}|v_m^{(+)}\rangle + \rho_m^{(-)}e^{i\beta_m(h/2-z)}|v_m^{(-)}\rangle\right]$. $|v_m^{(\pm)}\rangle$ denotes the transversal field of the $m^{th}$ waveguide mode supported by the corresponding slab with the infinity thickness ($h$=+∞), which propagates along $\pm z$ direction with the real propagation constant $\beta_m$. $N$ is the amount of the propagating waveguide modes. At the interface between the slab and free space ($z$=$h$/2), these propagating modes partly reflect back into the slab, simultaneously, partly transmit into free space, which can be described by an interface-reflection matrix $r$ ($N \times N$ version) and an interface-transmission matrix $t$ ($6 \times N$ version in the vicinity of **K** point). Taking in account the *z*-mirror symmetry of the slab, the column vector $\boldsymbol{\rho}^{(+)} = \left[\rho_1^{(+)}, \rho_2^{(+)}, ..., \rho_N^{(+)}\right]^T$ composed by the mode coefficients satisfies the eigen equation,

$$(fr)\boldsymbol{\rho}^{(+)} = \eta \boldsymbol{\rho}^{(+)}, \tag{2}$$

in which, $f = \mathbf{diag}\left[e^{i\beta_1 h}, e^{i\beta_2 h}, ..., e^{i\beta_N h}\right]$. The eigenvalue $\eta$ is equal to 1 and -1 for TE-like and TM-like modes of the slab, respectively. Equation (2) means the existence of BICs requires that the reflection matrix $r$ must be non-diagonal. Therefore, the inter-conversions among the waveguide modes in the slab (that is, each waveguide mode experiences not only reflection but also converts into other modes at the two interfaces) play an important role in the realization of BICs by the destructive interferences.

At the **K** point, due to the $C_{3v}$ symmetries of the PhCS [Fig. 1(a)], the matrix $r$ and $t$ are block-diagonal since only waveguide modes with the compatible symmetries can be interconverted and transmit into the radiation channels with the compatible symmetries at the interfaces. For the sake of simplicity, in our designed non-degenerate K-point BICs in Fig. 2, only two waveguide modes and one radiation channel are symmetry-matched and take part in the destructive interferences. While, the double-degenerate K-point BICs result from the destructive interferences dealing with four waveguide modes and two radiation channels. The previously reported anisotropy-induced BICs [5] and off-Γ BICs [10] belonged to the simplest scenario of the destructive interferences involving only two waveguide modes and one radiation channel. Where, one thickness (*h*)-independent necessary condition for the existence of BIC in the *z*-mirror-symmetrical slabs (for detailed derivation see the Supplementary Material Note 6) can be given by the matrix $r$ (reduced to $2 \times 2$ version) and $t$ (reduced to $1 \times 2$ version). That is,

$$\left|r_{11} - r_{12}\, t_{11}/t_{12}\right| = \left|r_{22} - r_{21}\, t_{12}/t_{11}\right| = 1. \tag{3}$$

By solving Eq. (2), we can obtain the thicknesses of PhCSs supporting BICs with desired symmetries and the far-field polarization states consistent with the results shown in Fig. 2 [for details see the Supplementary Material Notes 7-9]. Therefore, our analytical model could efficiently describe the realization of BICs formed by the destructive interferences and the far-field polarization diversity.

In conclusion, by constructing **K**-point BICs in honeycomb-lattices PhCSs, we have theoretically demonstrated that the far-field radiation close to BICs could exhibit the polarization diversity (e.g. circular polarization, elliptical and linear polarizations with variant orientations, polarization vortices with the different topological charges). The spatial symmetry and the interferences among the different radiation channels play important roles in determining the polarization state of the far-field radiation. The trajectories of the far-field polarization states on the Poincaré sphere provide a general tool to efficiently characterize the topological natures of BICs. Since *C*-points close to BICs are observed in this work, limiting to the closed loops enclosing BIC in the momentum space without containing any *C*-point, we define the topological charge carried by the BIC as the half of the winding number of the corresponding trajectory around the $S_3$ axis. It has been verified that the topological charge carried by the non-degenerate and double-degenerate BICs at K point are ±1 and 1/2, respectively. The polarization diversity close to BICs may bring about opportunities to generate and manipulate vector beams with diverse polarization states and different topological natures in the momentum space.

We are grateful to Professor Shuang Zhang and Dr. Biao Yang for fruitful discussions. The work was supported by National Science Foundation of China (11374367, 61307072, 61405056).

**Figures 1-4:**

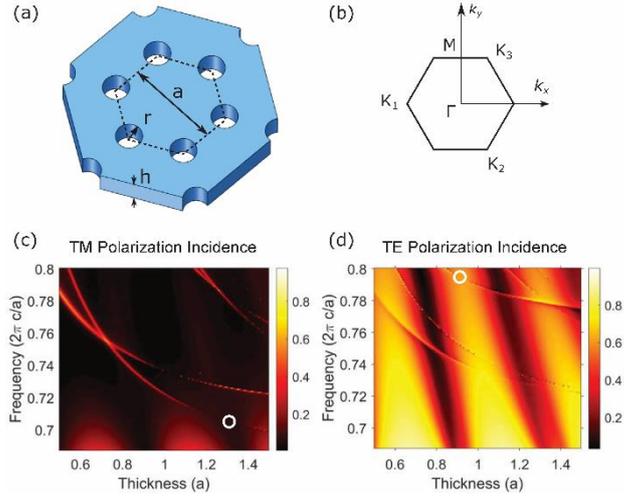

**FIG. 1** Schematic illustration of the two-dimensional photonic crystal slab with a honeycomb array of cylindrical holes etched in a free-standing dielectric slab (a). The first Brillouin zone (b) in the reciprocal space of the honeycomb lattice with positions of $\mathbf{K}_1=-K\mathbf{e}_x$, $K=4\pi/(3a)$. (c) and (d) Reflectivity spectra of TM and TE polarized plane waves with the wave vector matching with the $\mathbf{K}_1$ incident on the PhCS with different thicknesses, respectively. The two circles denote the positions where the two non-degenerate BICs appear.

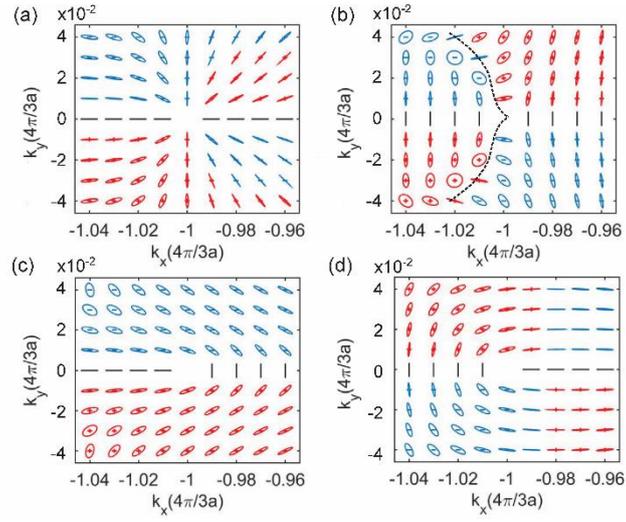

**FIG.2** Far-field polarization states of eigenmodes supported by the PhCSs in the momentum space close to the **K**-point BIC. (a), (b) Results of PhCSs with (*r*, *h*) equal to (0.15*a*, 1.27*a*) and (0.15*a*, 0.9074*a*) supporting a non-degenerate even BIC and odd BIC, respectively. (c)-(d) The low- and upper-band results of the PhCS with (*r*, *h*) equal to (0.143*a*, 1.036*a*) supporting one double-degenerate BICs. In (b), the ellipse (circle) denoted by +(-) means the polarization state of the far field is a right (left)-handed ellipse (circle).

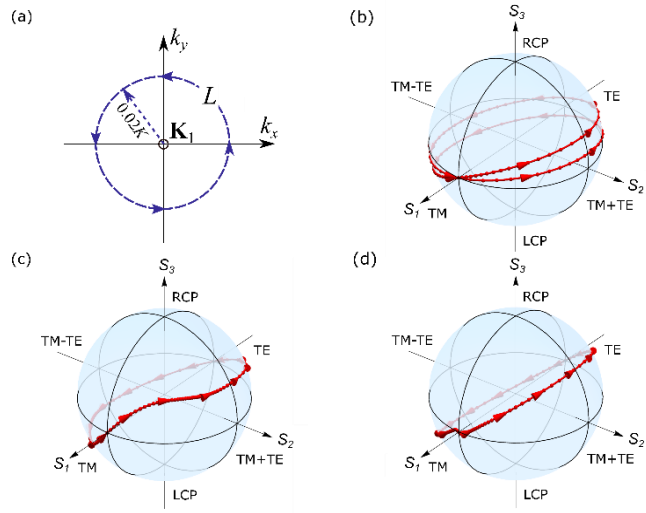

**FIG. 3** Trajectories of far-field polarization states of different PhCSs on the shell of the Poincaré sphere. (a) The selected anticlockwise circle centered at $\mathbf{K}_1$ point in the momentum space. (b) Trajectories of the PhCS supporting the non-degenerate even BIC [Figs. 2(a)]. (c) and (d) Trajectories of the PhCS supporting the double-degenerate BICs [Figs. 2(c) and 2(d)], respectively.

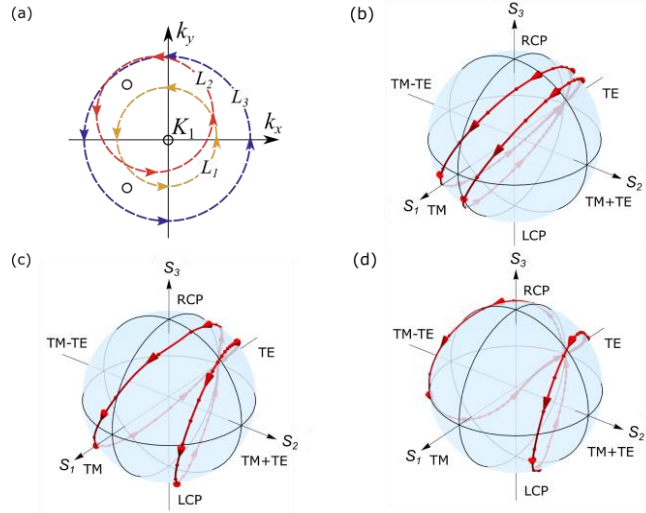

**FIG. 4** Trajectories of far-field polarization states of the PhCS supporting non-degenerate odd BICs [Fig. 2(b)] on the shell of the Poincaré sphere. (a) Three selected anticlockwise circles enclosing $\mathbf{K}_1$ point denoted by $L_1$, $L_2$, $L_3$. The two circles denote the positions of *C*-points. (b)-(d) Trajectories corresponding to the circles $L_1$, $L_2$, $L_3$, respectively.